\documentclass[twocolumn,showpacs,preprintnumbers,amsmath,amssymb]{revtex4}

\usepackage{graphicx}
\usepackage{dcolumn}
\usepackage{bm}

\begin{document}

\preprint{APS/123-QED}

\title{Diffusion of  spherical particles in microcavities}
\author{A. Imperio$^1$} \email{a.imperio@virgilio.it} %
\textbackslash\textbackslash 

\author{J.T. Padding$^2$} \email{j.t.padding@gmail.com}
\author{W.J. Briels$^1$}
 \homepage{http://cbp.tnw.utwente.nl} 
\affiliation{$^1$ Computational Biophysics, University of
Twente, P.O. Box 217, 7500 AE, The Netherlands\\
$^2$ IMCN, Universit\'e catholique de Louvain, Croix du Sud 1,
Louvain-la-Neuve, Belgium
}%

\date{\today}

\begin{abstract}
The diffusive motion of a colloidal particle trapped inside a small cavity
filled with fluid is reduced by hydrodynamic interactions with 
the confining walls. In this work, we study these wall 
effects on a spherical particle entrapped in a closed cylinder. We
calculate the diffusion coefficient 
along the radial, azimuthal and axial direction
for different
particle positions. At all locations the diffusion is smaller than in
a bulk fluid and it becomes anisotropic near the container's walls. 
We present a simple model which reasonably well desribes the
simulation results for the given dimensions of the cylinder, which are
taken from recent experimental work.
 
\end{abstract}

\pacs{Valid PACS appear here}
\maketitle

\section{\label{sec1}Introduction}
Modeling the motion of colloidal particles under confinement is
of interest for many industrial applications which involve
the entrapment of particles in microporous materials  and diffusion controlled
reaction mechanisms. An important class is that of 
 coating processes where the particles have to diffuse towards a wall
 before being
immobilized. Other applications involve microfluidic chips
\cite{vanommering09,vanommering06} and surface-based biosensor
\cite{squires08}.
 Furthermore the confinement of particles in more than one dimension
 is relevant for the synthesis of colloids inside droplets
 \cite{shestopalov04}
and for understanding the transport of proteins, lipid granules
and other macromolecular assemblies through a biological cell.

Most of the past work on modeling the mobility of a particle in confinement
has focussed on a particle in proximity of a flat wall or within an
infinitely long cylinder \cite{happel91,kim93,goldman67,falade88,hirschfeld84}.
The motion of an arbitrarily shaped particle near boundaries has been
studied in \cite{cox67}, while the settling of a particle inside a
cylinder closed at the bottom but characterized by a free surface at
the top has been analyzed in \cite{sonshine66}. Finally for a
historical perspective
on the motion of a sphere in an uncompressible fluid, from Stokes' law to
Fax\'en's solution of the Navier-Stokes equations for particular bounded
systems, see also \cite{lindgren99}. Under the assumption of creeping
motion, previous investigations are
mainly based on various approximate perturbative schemes, which are
generally not valid when the size of the particle, the size of the
cavity and the gap between particle and wall are all comparable, which is
the case we are interested in.
Recently, however, a semi-analytical procedure for the calculation of
the mobility matrix of a spherical particle inside a cylinder has
been presented in \cite{bhattacharya10}, which can be used under a
wide range of confinement conditions.
 
The motion of a particle in a closed three dimensional container has
been studied less intensely. The flow field around a sphere has been studied in \cite{blake79,sano87} in various confined geometries.
The motion of a particle along the axis of closed cylinders and closed cones is discussed in ref. \cite{lecoq07}, where 
experimental and theoretical results are presented for a millimeter sized sphere
settling in a very viscous silicone oil. Agreement is found until the wall-particle separation is of the
order of the particle diameter. Recently the diffusion coefficients of
$\mu m$ sized spheres moving in a squat cylinder containing a water
solution has been studied \cite{eral10}, showing that the diffusion
is strongly affected by the confinement and substantial corrections to
the Stokes' law are required.

The analysis of the motion of a particle inside a cavity can be
based upon the solution of Navier-Stokes equations under particular
regimes, such as the creeping motion. However, such equations are
not always easily solved, especially for complex geometries or when
there are many particles in suspension. Moreover they are based on
the hypothesis that the solvent is a continuum. But in
conditions relevant for microfluidic devices, this hypothesis is not
always fulfilled and fluctuations of the solvent velocities may be
relevant. 
To study diffusion of colloids in microcavities it is therefore important to test different
tools, like simulations in which the solvent is explicitly introduced, or in which the solvent velocity is allowed to fluctuate.
In this work we use the Multi-Particle Collision Dynamics technique,
initially proposed by Malevanets and Kapral \cite{malevanets99}. 
It consists of a coarse-grained model to describe the
solvent completed with specific rules (see below) to describe
solvent-solute interactions. It has been successfully used to
study hydrodynamic interactions in complex fluids, at equilibrium
and far from it. For a review, see \cite{gompper09}.

We will briefly introduce the simulation technique in section \ref{sec2}. 
In section \ref{sec3a} we discuss the measurements of friction and diffusivity.
In section \ref{sec3} we will analyze the simulations results for the diffusion of a spherical particle
inside a microcavity in the shape of a closed cylinder. We will show that
the friction is different along the radial, azimuthal and axial directions,
and that the corresponding diffusion coefficients change in different ways with the particle's radial and axial position.
In section \ref{sec4} we will compare our results with experimental and theoretical data.
A simplified model to interpret our results is given in section \ref{sec5}.
Finally, we give our conclusions in section \ref{sec6}.

\section{\label{sec2}Simulation technique}

A severe difficulty in simulating colloidal solutions is to correctly deal with the large range of time and
 length scales involved. Whereas the solvent velocities generally relax in less than a picosecond, colloids
 typically diffuse over a length equal to their own size in seconds. Fortunately it is often not necessary
 to exactly reproduce such time and length scales. A coarse-grained approach is 
 possible in which the scales are telescoped together \cite{padding06},
 while keeping an order of magnitude difference between each relevant time scale.
 
As already mentioned, we will use Multi-Particle Collision Dynamics
(MPCD) \cite{malevanets99,gompper09} to describe the solvent degrees
of freedom, with no-slip boundary conditions for the solvent-solute
interactions. The implementation of the no-slip 
boundary conditions has been discussed in detail in \cite{imperio11}. The dynamics of the system is
made up of two steps: streaming and collision. In the streaming step,
the position and  velocity of each particle is propagated for a time
$\delta t$ by solving
 Newton's equations of motion. During the collision a lattice
 is placed in the system \cite{ihle01}. This lattice is
 typically cubic, with a lattice constant $a$. Solvent particles are
 then attributed to the cells in which they happen to be positioned. Next the velocities $\boldsymbol{v}_i$ of all particles are stochastically
 rotated, relative to the center of mass
 of the corresponding cells according to the formula:
\begin{equation}
\boldsymbol{v}'_i(t)=\boldsymbol{u}+\boldsymbol{\Omega}
\{\boldsymbol{v}_i(t)-\boldsymbol{u}\},
\label{eq.col1}
\end{equation}
where $\boldsymbol{u}$ is the mean velocity of the particles within the collision cell to which particle $i$ belongs
and $\boldsymbol{\Omega}$ is a matrix which
rotates velocities by a fixed angle $\alpha$ around a randomly
oriented axis.
Through the stochastic rotation of the velocities, the solvent
particles can effictively exchange momentum without the need of introducing direct forces
 between them. As the collision step preserves
 mass, linear momentum and energy, the correct hydrodynamic
behavior is obtained on the  mesoscopic scale
\cite{malevanets99,malevanets00}.

When colloids are present, Newton's
equations of motion are solved also for them during the streaming
step. However special care must
be taken when no-slip boundary conditions are applied
on the colloidal surface and on the confining walls. The
implementation of the no-slip boundary conditions is described in the
following.\\ 
\protect{\em{Streaming step:}} when a solvent particle
crosses the colloid (or wall) surface, it is moved back to the
impact position. Then a new velocity is extracted from the
following distributions for the tangent ($v_t$) and the normal
component ($v_n$) of the velocity, with respect to the surface velocity:
\begin{equation}
p(v_n)=\frac{m v_n}{k_BT} \exp\left(-\frac{mv_n^2}{2k_BT}\right),\label{eq.vn}
\end{equation}
\begin{equation}
p(v_t)=\sqrt{\frac{m}{2\pi k_BT}} \exp\left(-\frac{mv_t^2}{2k_BT}\right)\label{eq.vt}.
\end{equation}
Here $m$ is the mass of the solvent particle, $k_B$ the Boltzmann's constant, and $T$ the
temperature of the system.
Once the velocity has been updated, the particle is displaced for the remaining
part of the integration time step.\\
\protect{\em{Collision step:}} virtual particles (VP) are inserted
  in those parts of the collision cells which are physically occupied by
  the colloid or by the container walls. The VP density matches that of the MPCD solvent,
  while their velocities $\boldsymbol{v}^{VP}_i$ are obtained from a Maxwell-Boltzmann
  distribution, whose average velocity is equal to the local velocity of the
  colloid or to the velocity of the walls, and the temperature is the
  same as in the solvent. Notice that we do not take rotations of the particles into account. VP are sorted into the collision cells according to their coordinates, exactly as the MPCD particles.
 This method is similar to that used in
 \cite{lamura01,winkler09,gotze07,downtown09}.
Then the average velocity of
  the center of mass of the cell is computed as:
\begin{equation}
\boldsymbol{u}=\frac{
              \overset{n_{MPCD}}{\sum_{i=1}}\boldsymbol{v}_i(t)+
              \overset{n_{VP}}{\sum_{i=1}}\boldsymbol{v}^{VP}_i(t)
}{n_{MPCD}+n_{VP}},
\label{eq.newv}
\end{equation}
where $n_{MPCD}$ and $n_{VP}$ are the number of MPCD particles and VP, respectively, belonging to the same cell.
Velocities of both MPCD and VP  belonging to the same cell are rotated according to the rule given in
  eq.(\ref{eq.col1}). 

Due to the exchange of momentum between the solvent and the colloidal particle,
 the force exerted upon the latter can be expressed as
 $\bf{F}=\bf{f}_s+\bf{f}_c$, where
 $\bf{f}_s$ and 
 $\bf{f}_c$ are the forces during the streaming and the
 collision step, respectively. The former can be calculated as: 
\begin{equation}
\protect{\bf{f}}_{s}=-\frac{1}{\delta t}\overset{Q}{\sum_{i=1}}
\Delta \protect{\bf{p}}_i,
\label{eq.forces}
\end{equation}
where $Q$ is the number of MPCD particles which have
crossed  the surface of the colloid between two
collision steps, and $\Delta \protect{\bf{p}}_i = m\Delta \boldsymbol{v}_i$ is the change in momentum
of solvent particle $i$, which has been scattered by the colloid.
The force exerted during the collision step is:
\begin{equation}
\protect{\bf{f}}_{c}=\frac{1}{\delta t}\overset{q}{\sum_{i=1}}\Delta
\protect{\bf{p}}_i^{VP},
\label{eq.forcec}
\end{equation}
where $q$ is the total number of virtual particles which belong to a
tagged colloid, and $\Delta \protect{\bf{p}}^{VP}_i$ is the change in momentum of
the virtual particle $i$ as a consequence of the collision step.

The force exerted by the solvent on the colloidal particle, will be
used in section \ref{sec3} in order to measure the friction on the particle
itself. 

In our simulations we use a collision angle $\alpha = \pi/2$, an average solvent particle density
of 5 particles per collision cell, and a collision time step $\delta t = 0.1 t_0$, where $t_0 = a\sqrt{m/k_BT}$
is the simulation unit of time. This choice leads to a mean-free path which is an order of magnitude smaller
than the collision cell size $a$, which is important to simulate the dynamics of a colloid in a liquid instead
of a gas \cite{padding06}.
Simulations have been performed with a colloidal sphere of radius $R_{col}=3a$,
which is sufficiently large to accurately resolve the hydrodynamic field to
distances as small as $R_{col}/6$, as already shown in Refs. \cite{padding06,padding10}.
The radius and height of the closed cylinder are $R_{cyl}=H_{cyl}=54a$. In this way we can make
a direct comparison with the experimental case discussed in
\cite{eral10}. The typical length of a
simulation run is about $1.5 \cdot 10^6$ integration time steps. 

\section{\label{sec3a}Measuring friction and diffusivity}

\begin{figure}
\includegraphics[width=5cm,height=4cm]{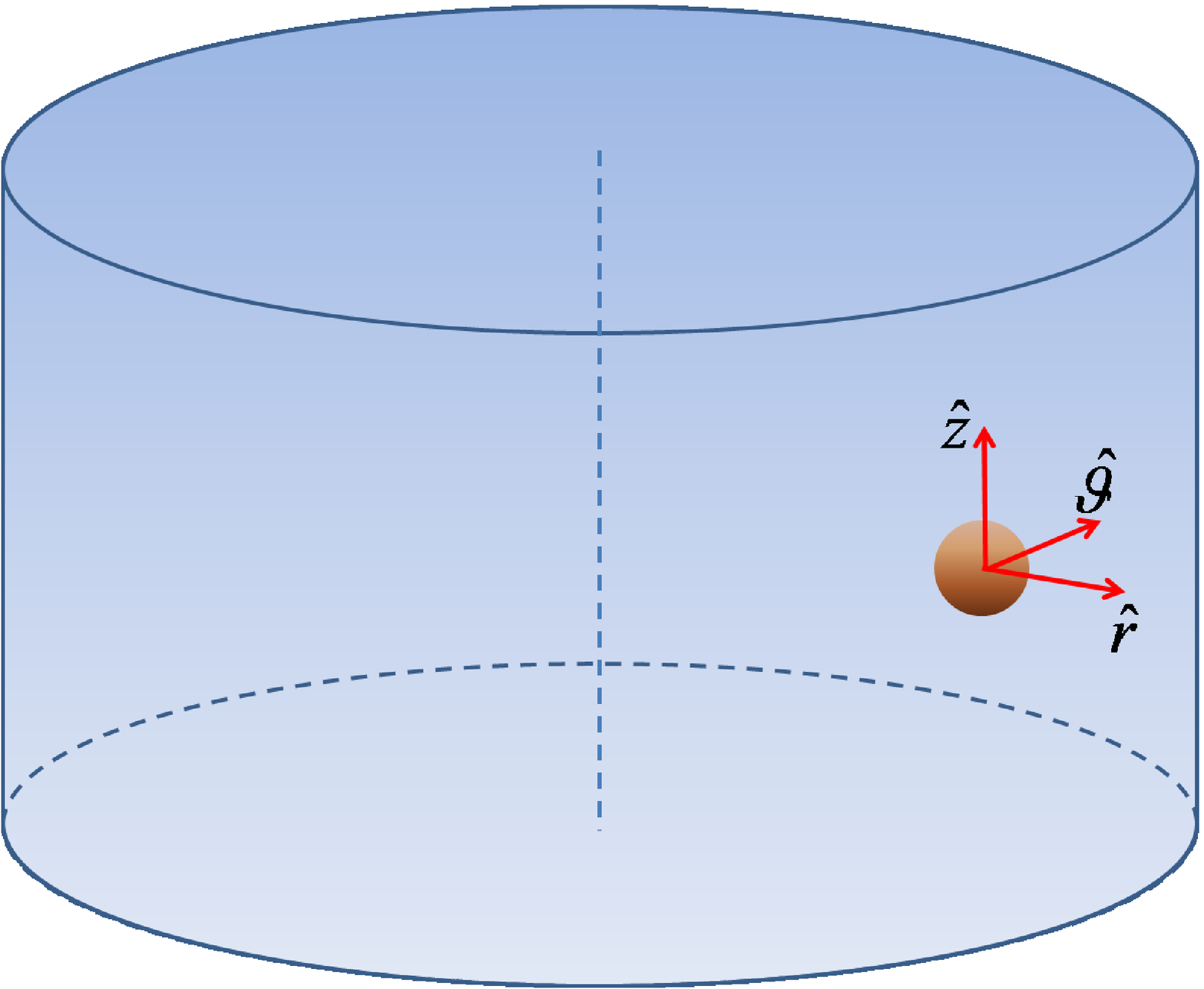}\\
\vspace{0.5cm}
\includegraphics[width=5cm,height=4cm]{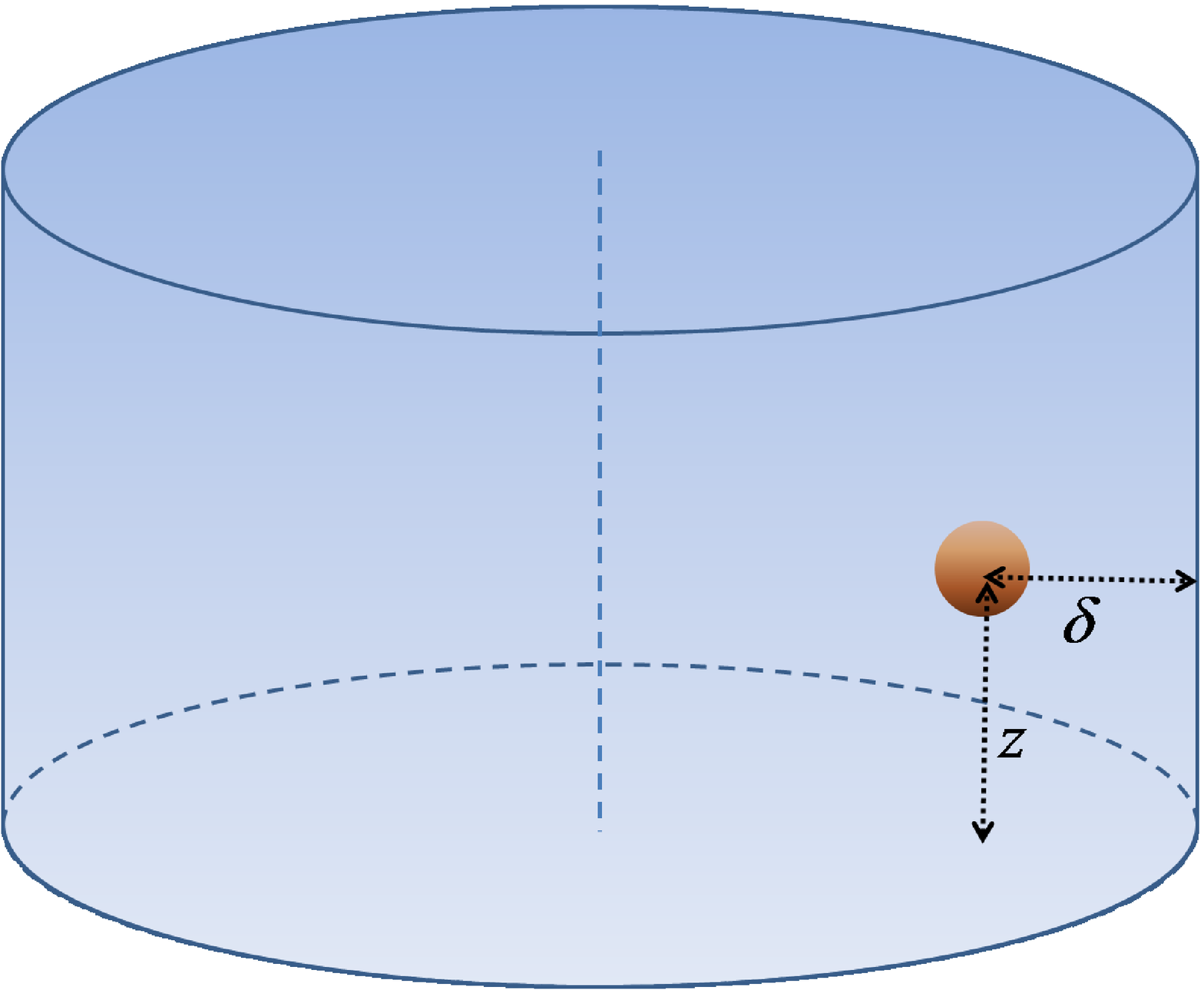}\\
\caption{\label{fig:geometry1}\footnotesize{ 
For a generic sphere position, we plot the unit vectors of the
radial ($\hat{r}$), azimuthal ($\hat{\theta}$) and axial ($\hat{z}$)
directions. The distance from the
cylindrical walls is $\delta$, while the distance from the bottom wall
is $z$. When the
sphere is in the mid-plane $z~=~H_{cyl}~/~2$, when the sphere is
placed along the axis $\delta~=~R_{cyl}$.
}}
\end{figure}

The drag force $\mathbf{F}$ and torque $\mathbf{T}$ acting on a particle that moves through
a fluid with a steady velocity $\mathbf{V}$ and steady angular velocity $\mathbf{\Omega}$ can be expressed as
\begin{equation}
- \begin{pmatrix}
\mathbf{F}\\
\mathbf{T}
\end{pmatrix}
=\,\mathbf{M}
\begin{pmatrix}
\mathbf{V}\\
\mathbf{\Omega}
\end{pmatrix},
\label{eq.frictiontot}
\end{equation}
where $\mathbf{M}$ is the friction tensor represented by a $6 \times 6$ matrix. The friction matrix  can be subdived in 4 blocks,
\begin{equation}
\mathbf{M} =
\begin{pmatrix}
\Xi^t&\Xi^{tr}\\
\Xi^{rt}&\Xi^r
\end{pmatrix},
\label{eq.tensor}
\end{equation}
where the superscript $t$ stands for translation, $r$ for rotation and $tr$
and $rt$ for the coupling between rotational motion and (translational) force and
between translational motion and (rotational) torque, respectively.
The blocks of the matrix $\mathbf{M}$ can be obtained in the following manner. 
During the simulation the colloidal sphere is kept in a fixed position and fixed orientation by applying an
external constraint force
$\mathbf{F}^c$ and external constraint torque $\mathbf{T}^c$. The constraint
force and torque are equal, but opposite, to the total force and torque
exerted by the solvent on the sphere
  $\mathbf{F}^c=-\mathbf{F}$ and $\mathbf{T}^c=-\mathbf{T}$.
Through the fluctuation-dissipation theorem the autocorrelation
functions of $\mathbf{F}^c$ and $\mathbf{T}^c$ can be connected to each of the components of
the friction tensor. For example the explicit expression for
 $\boldsymbol{\Xi}^t$ \cite{akkermans00,padding10} is:
\begin{equation}
\Xi_{\alpha\beta}^t=\frac{1}{k_BT}\int_0^{\infty} \mathrm{d} \tau \left\langle F_{\alpha}^c
(\tau_0+\tau) F_{\beta}^c(\tau_0) \right\rangle_{\tau_0}.
\label{eq.friction}
\end{equation}
Similarly we can also measure:
\begin{equation}
\Xi_{\alpha\beta}^{tr}=\frac{1}{k_BT}\int_0^{\infty} \mathrm{d} \tau \left\langle F_{\alpha}^c
(\tau_0+\tau) T_{\beta}^c(\tau_0) \right\rangle_{\tau_0}.
\label{eq.coupling}
\end{equation}
and so on. The quantities
 $\alpha,\beta$ refer to two of the cylindrical coordinates
$r,\theta,z$ and the average is taken over many different time origins $\tau_0$.
As shown in Fig.~(\ref{fig:geometry1}), for a given position of the sphere we can define an orthogonal frame with unit vectors pointing in the local $r$, $\theta$ and $z$ directions. Because the sphere is in a fixed position, this orthogonal frame does not change during the simulations.

\begin{figure}
\includegraphics[width=7cm,height=8cm,angle=270]{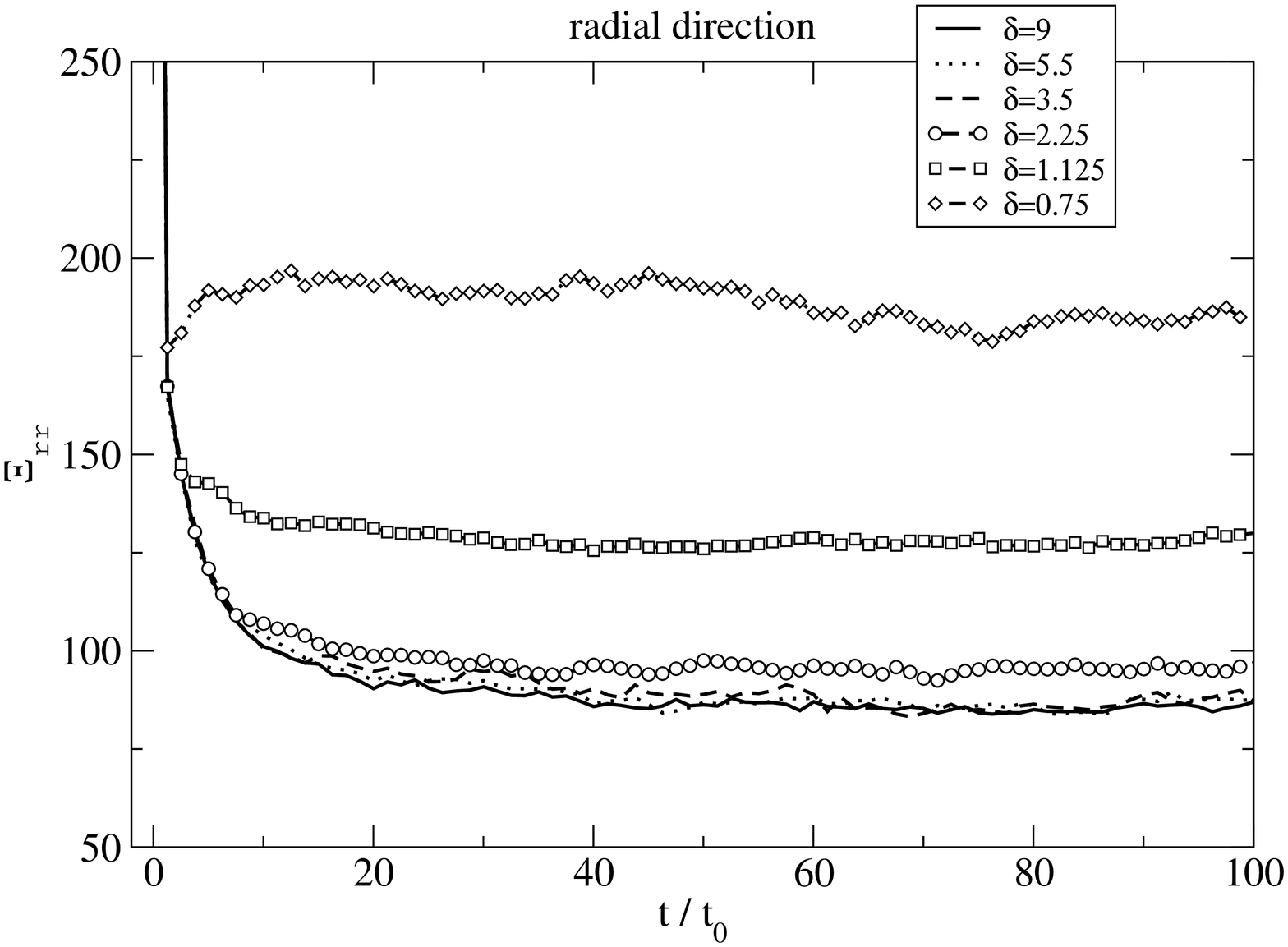}\\
\vspace{-0.5cm}
\includegraphics[width=7cm,height=8cm,angle=270]{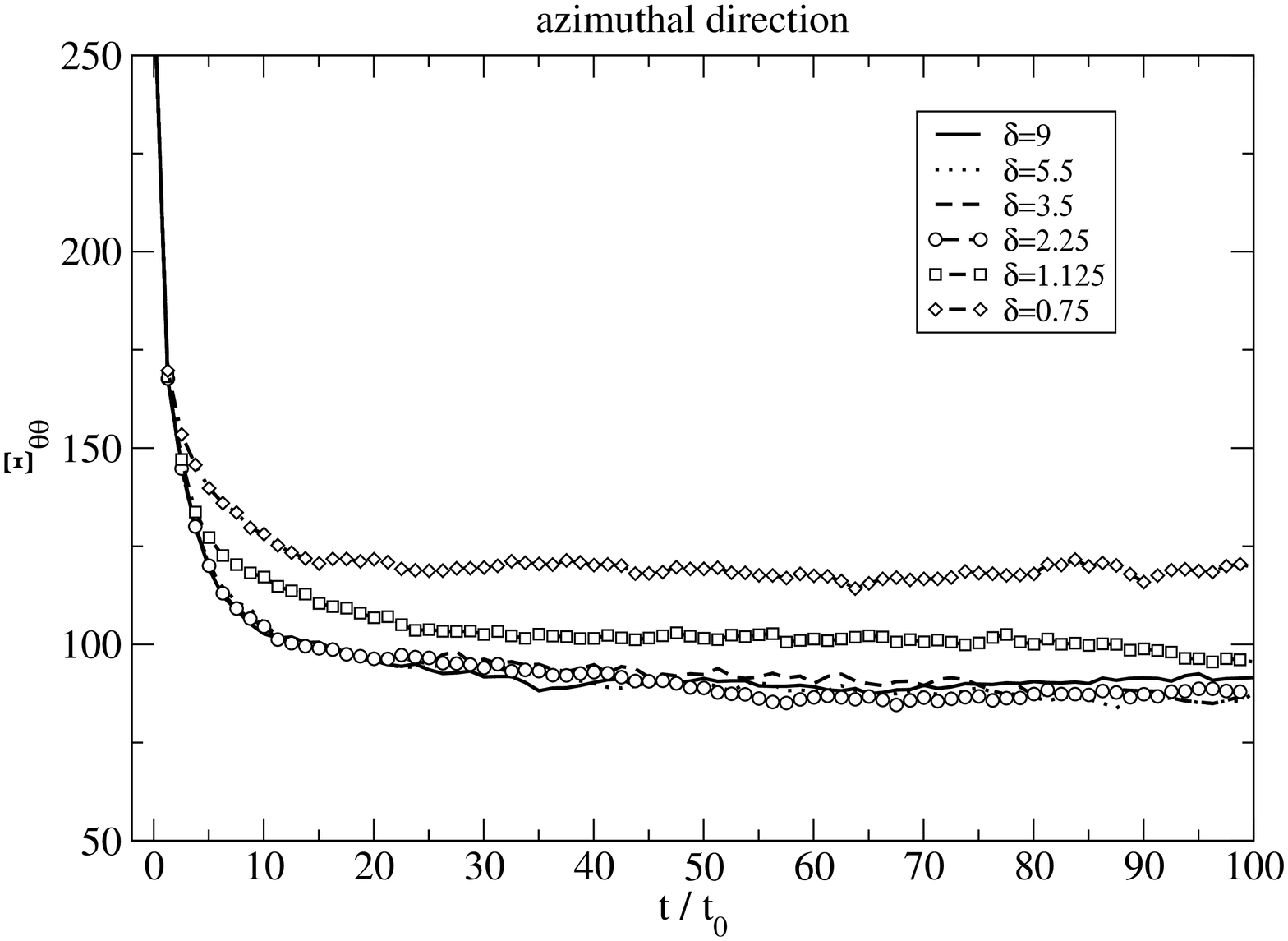}
\caption{\label{fig:autocor}\footnotesize{ Running integral 
of the autocorrelation
  function of the constraint force $\protect{F}^c$ applied to the sphere in the
  mid-plane. The distance  $\delta$ from the cylindrical wall is in
  units of the particle diameter. Time is in units of
  $t_0=a(m/k_BT)^{1/2}$ and friction in units of $m/t_0$. Upper panel: radial friction. Bottom panel:
  azimuthal friction, which is very similar to the axial friction (not
  shown in this picture).
 }}
\end{figure}

\begin{figure}
\includegraphics[width=7cm,height=8cm,angle=270]{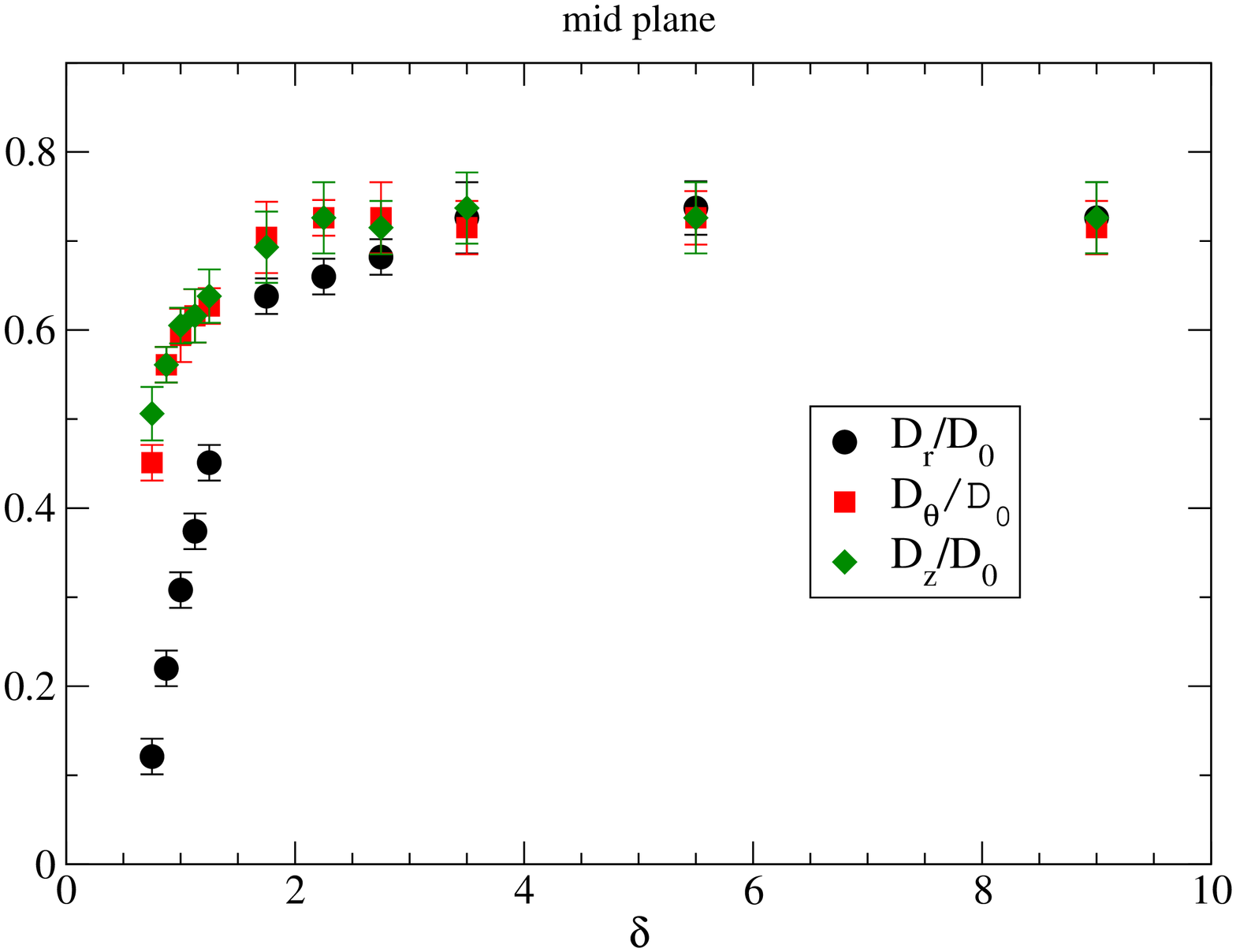}\\
\includegraphics[width=7cm,height=8cm,angle=270]{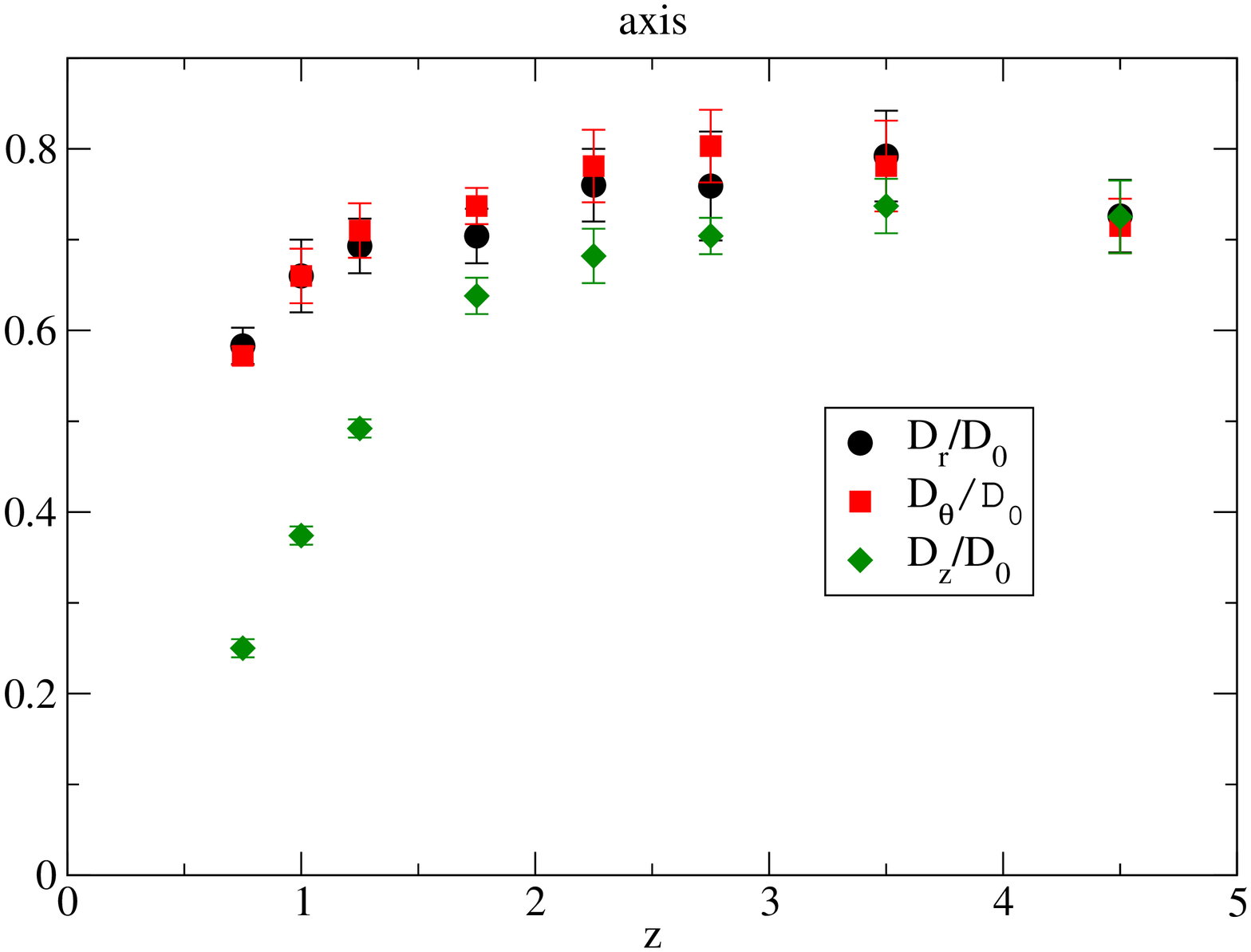}
\caption{\label{fig:diffusion}\footnotesize{Diffusion coefficients
    along the radial ($\hat{r}$),
    azimuthal ($\hat{\theta}$) and axial ($\hat{z}$) direction. 
Upper panel: the
    sphere is in the mid-plane and $\delta$ is the separation from
    the cylindrical walls. Bottom panel: the sphere is placed along
    the cylinder axis and $z$ is the distance from the closest
    end-wall.
     Distances are in units of
    the particle diameter. Diffusion is in units of the bulk value $D_0$ (see main text).
    }}
\end{figure}

The friction tensor $\mathbf{M}$ is connected to the diffusion tensor
through the Einstein relation $\mathbf{D}=k_BT
\boldsymbol{M}^{-1}$. The diffusion tensor can be subdived in blocks
too:
\begin{equation}
\mathbf{D}=
\begin{pmatrix}
D^t&D^{tr}\\
D^{rt}&D^r
\end{pmatrix},
\label{eq.diffusion}
\end{equation}
where $D^t$ is the translational diffusion tensor,
$D^r$ the rotational diffusion tensor, and $D^{tr}$ and $D^{rt}$ determine the amount of
cross-correlation between translational and rotational displacements.

A strong coupling between translation and rotation is expected when the
particle gets very close to a wall. This will cause a torque-free particle to rotate when it is
translated along the wall,  and vice versa a force-free particle to translate when it is rotated near a wall. 
However, after a preliminary analysis on the walls' effects, based upon a sphere nearby
a flat wall \cite{goldman67} and inside an infinitely long cylinder
\cite{falade88}, for the particular gaps we are interested
in, we find that such a coupling is very
small: in the worst case scenario ($d/R=0.5$) the effect of including translation-rotation coupling on the
translational diffusion is less than $2\%$. 
We expect that the contribution due to the coupling between rotation and translation
is small in the closed cylinder as well, so that we can neglect the terms $\Xi^{tr}$ and $\Xi^{rt}$ in the
friction tensor.

In this paper we focus on two cases: one in which the particle is set
in the mid-plane of the cylinder and another in which it is located along its axis. For both these cases,
symmetry requires that $\Xi^t$ is diagonal.  Therefore the
translational diffusion tensor $D^t$ is also diagonal and, under the
hypothesis that the coupling between rotation and translation is
negligible, we set
 $D^t_{rr}= k_BT/\Xi^t_{rr}$, $D^t_{\theta\theta}= k_BT/\Xi^t_{\theta\theta}$ and
$D^t_z=k_BT/\Xi^t_{zz}$.

Because in this paper we measure only the translational diffusion and not
the rotational diffusion, we drop the superscript $t$ and  simplify the notation as $D_r$, $D_\theta$ and $D_z$.
Some specific components of the running integral of the autocorrelation
function of $\mathbf{F}^c$  are shown in Fig.~(\ref{fig:autocor}):
the sphere is in the mid-plane but at different distances
 $\delta$ from the cylindrical walls. 
In particular in the upper panel we show the radial component of the
friction $\Xi_{rr}$, while in the bottom panel the azimuthal one
$\Xi_{\theta\theta}$. 

Finally we note a peculiarity which is linked to the fact that in our simulations a colloidal particle collides with only a 
few solvent particles within each particular time step $\delta t$.
If we consider a particular direction $\alpha$,  there are essentially
two contributions to the total friction:
one coming from the local Brownian collisions with the particles of the fluid ($\xi^B$),
while the other is due to the long-range hydrodynamic interactions (flow fields) ($\xi^H$). 
A simple empirical formula says that the hydrodynamic and the Brownian friction should
be added in parallel \cite{hynes77,lee04} in order to obtain the total friction:
\begin{equation} 
1/\Xi_{\alpha\alpha}=1/\xi^B+1/\xi^H
_{\alpha\alpha}
\label{vale}
\end{equation}
$\xi^B$ does not depend on the sphere's position and corresponds to the height of the
short-time peak of the integral in eq.~(\ref{eq.friction}); its value is $645\pm 1$ (in units $m/t_0$) for all
diagonal components.
The long-time limit of the profiles shown in Fig.~(\ref{fig:autocor}) provides the total
friction upon the sphere.
Thus, the hydrodynamic term can be extracted by  inverting
eq.(\ref{vale}). For validation of this method and further
 applications of this concept see \cite{padding06,padding10}.
 
 \section{\label{sec3}Simulation results}
 
The measured diffusion coefficients are
 summarized in Fig.~(\ref{fig:diffusion}) for the mid-plane (upper
 panel) and for the cylinder axis (bottom panel). The diffusion
 is normalized to the bulk value $D_0$ for a sphere with no-slip
 boundary conditions upon its surface. The latter is provided by the
 Stokes-Einstein law: $D_0=k_BT/(6\pi \nu R_{col})$, where $\nu$ is the solvent viscosity.
 
First we focus on the case where the sphere is located in the mid-plane of the cavity. 
When the colloidal sphere is set near the center of this mid-plane (large $\delta$), we expect that
$D_r=D_\theta$ because of symmetry. Actually 
we observe that all three diagonal components are almost the
same and their value is significantly lower ($25\%$) than in a bulk system.
When the sphere is shifted towards the cylindrical wall, the
diffusion decreases even more. For off-centered positions, the radial component $D_r$ decreases more rapidly
than $D_\theta$ and $D_z$ do.

Next we focus on the case where the sphere is located on the central axis of the cavity, at
a distance $z$ from the nearest end-wall.
Although in this case technically no azimuthal direction can be defined, we have made
two orthogonal measurements perpendicular to the $z$-axis, and labeled them as $D_r$ and $D_\theta$
in the bottom panel of Fig.~(\ref{fig:diffusion}). As expected, because of symmetry we find
$D_r=D_\theta$ within uncertainties. 
Close to the end-wall, $D_z$ decreases much faster than the other two
components do. This situation is similar to the case of a sphere close to
a flat wall. However the similarity is only qualitative and not
quantitative, as we will discuss in section \ref{sec5}.

\section{\label{sec4}Comparison with experimental and theoretical predictions for the closed cylinder}

First we compare our data with some theoretical results published recently and next with some recent experimental results.
It is interesting to notice that our results presented in the upper panel of Fig.~(\ref{fig:diffusion}) are in agreement with the findings in \textit{et al.} \cite{bhattacharya10}, where hydrodynamic calculations are performed of the drag on a particle in 
 an infinitely long cylinder. In particular
the authors observe that for $R_{cyl}\ge 4R_{col}$, all three diagonal
components of the friction increase monotonically when the particle
approaches the wall.  The radial friction is expected to vary
stronger than the other two frictions, because its motion is perpendicular to the confining
walls, while in the other cases the motion is tangent to the
wall. As discussed in \cite{bhattacharya10},  the resistance for the
radial motion is inversely proportional
to the gap between particle and wall, while for the tangential motions
it varies logarithmically with the gap. Qualitatively our results are
similar, in the sense that the azimuthal diffusion for small
particle-wall separations  varies slower than the radial diffusion does;
this suggests that the role of the cylindrical wall is dominant
in this region. However, we will show that for our squat cylinder the role of the end walls
cannot be neglected at any sphere position (section \ref{sec5}).

We have performed  two additional simulations in order to compare the results with the theoretical predictions
in \cite{lecoq07} for the axial friction $\gamma_z=\xi_z/\xi_0$ in a closed cylinder.
Here $\xi_0=6\pi \nu R_{col}$
corresponds to the friction on a sphere in an unbounded system.
The two cases that we have chosen are, one in which the cylinder is very narrow
and long, and one in which it is squat. The results are summarized
in Table~\ref{tab:lecoq}. The agreement is good.
\begin{table}
\caption{\label{tab:lecoq}\footnotesize{ Axial friction upon a sphere
    in a closed
 cylinder. The radius and the height of the cylinder are in units of
 the sphere radius. Theoretical predictions corresponds to
 \cite{lecoq07}.
The friction is normalized to the bulk value which is $\xi_0=6\pi\nu R_{col}$.
 }}
\vspace{0.5cm}
\begin{ruledtabular}
\begin{tabular}{rrcr}
$R_{cyl}$& $H_{cyl}$ & Lecoq 2007 & simulation\\
\hline
4 & 20 & 1.979 & 2.03 $\pm$ 0.08 \\
10.7 & 6 & 1.843 & 1.82 $\pm$ 0.07 \\
\end{tabular}
\end{ruledtabular}
\end{table}

In case the closed cylinder is very long,
 the friction on a sphere in the mid-plane should not be very different from that
predicted for a sphere in an infinitely long tube. For example, for the case
$R_{cyl}=4 R_{col}$ the friction inside an infinitely long cylinder
can be found in \cite{bhattacharya10}. In particular for a sphere
placed on the cylinder axis, the values are: $\gamma_r=1.763$ and
$\gamma_z=1.992$. Our simulation results for a closed
cylinder with the same radius but finite length $H_{cyl}=20 R_{col}$ agree within error bars:
$\gamma_r=1.79\pm0.07$ and $\gamma_z=2.03\pm0.08$. We can 
conclude that, for a long cylinder the role of the end walls is minimal.

\begin{figure}
\includegraphics[width=7cm,height=8cm,angle=270]{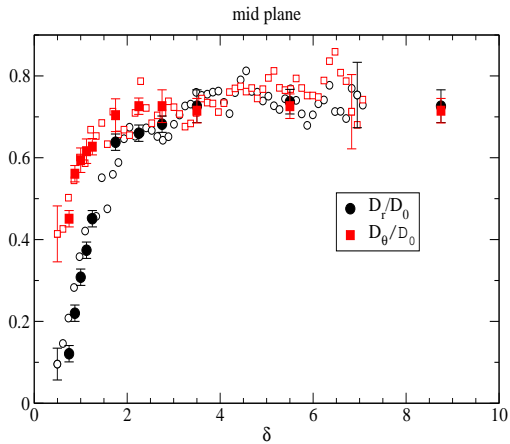}
\caption{\label{fig:experiments}\footnotesize{ A comparison between simulation
    results (filled symbols) and experimental data \cite{eral10}
    (open symbols)
    for the diffusion coefficient in the mid plane of a closed
    cylinder. Circles: radial diffusion. Squares: azimuthal
    diffusion. The wall-particle distance $\delta$ is in units of the
    particle diameter $d$. $R_{cyl}=9d$ and $H_{cyl}=9d$. For sake
    of clarity, we show the
    error bars for only a few experimental data points.
    }}
\end{figure}

The dependence of the diffusion on the spatial position  of a
particle inside a closed cylinder has been recently studied through experiments in
\cite{eral10}. The authors discuss the case of a $\mu m$ size
particle in an aqueous suspension. The microcavity is produced via
soft lithography where PDMS membranes are used. In
Fig.~(\ref{fig:experiments}) we
compare their experimental results (open symbols) with our simulation results (filled symbols) in the mid-plane for
the radial and the azimuthal components. The setup of the experiments
did not allow the measurement of the axial diffusion. The agreement
between the two sets of data is good. In particular the region where
$D_r$ and $D_\theta$ begin to differ is well captured. 

The good agreement with both experimental and theoretical predictions
reaffirms that MPCD simulations enable us to predict quantitatively
the friction on a colloidal particle in a confined space.

\section{\label{sec5}Discussion of wall effects}
We will now discuss the effects of confinement on the diffusivity of a colloidal particle, by
focussing on the effects of each individual type of boundary, i.e. cylinder or flat wall.

Generally speaking, for small Reynolds numbers, the correction term to
the friction on a confined particle is
$\gamma~=~\xi /\xi_0$, where $\xi_0$ is the friction on a sphere in an
unbounded solvent.
In principle $\gamma$ depend on the direction of motion.
Since there are no sufficiently general solutions available of
the Navier-Stokes equations for a closed cylinder, we
will try to estimate the total wall corrective factor as a combination
of the corrective terms due to simplified geometries, such as the
infinitely long cylinder, flat walls and square ducts. 
For ease of reference we reproduce explicit formulas available in the literature.

The wall correction for the axial friction on a sphere at an eccentric position is, approximately, \cite{happel91}:
\begin{equation}
\gamma_{z}^{cyl}=1+\left(2.10444-0.6977
\left(\frac{R_{cyl}-\delta}{R_{cyl}}\right)^2\right)
\left(\frac{R_{col}}{R_{cyl}}\right).
\label{eq:axcylinder}
\end{equation}
This expression is valid mainly near the axis of the cylinder. 
If the sphere is exactly on the axis, a more precise correction
due to Bohlin \cite{bohlin60}, also reported in \cite{happel91}, is:
\begin{eqnarray}
\gamma_{z}^{cyl}=\left[ 1-2.10444\chi+2.08877\chi^3-6.94813\chi^5+ \right. \nonumber \\
 \left.-1.372\chi^6+3.87\chi^8-4.19\chi^{10}\right]^{-1},
\label{eq:percylinder}
\end{eqnarray}
where $\chi=R_{col}/R_{cyl}$.
When the sphere is small (relative to the cylinder's diameter)
and very close to the cylinder wall, it is expected that the
curvature of the wall
is not very important and that the sphere behaves like near to a flat
wall \cite{brenner88}. In such a case the wall correction for the
parallel motion is \cite{happel91}:
\begin{eqnarray}
\gamma_{||}^{wall}=\left[1-\frac{9}{16}\left(\frac{R_{col}}{h}\right)+
\frac{1}{8}\left(\frac{R_{col}}{h}\right)^3+\right. \nonumber \\
\left.
-\frac{45}{256}\left(\frac{R_{col}}{h}\right)^4-\frac{1}{6}
\left(\frac{R_{col}}{h}\right)^5\right]^{-1},
\label{eq:parallel}
\end{eqnarray}
while for the perpendicular motion it is:
\begin{eqnarray}
\gamma_{\perp}^{wall} &=&
\frac{4}{3}\sinh(\alpha)\sum_{_n=1}^{\infty}\frac{n(n+1}{(2n-1)(2n+3)}\times \nonumber \\
&& \left[
\frac{2\sinh(2n+1)\alpha+(2n+1)\sinh2\alpha}{4\sinh^2(n+0.5)
\alpha-(2n+1)^2\sinh^2(\alpha)}-1
\right]. \nonumber \\
\label{eq:perpendicular}
\end{eqnarray}
Here $\alpha=\cosh^{-1}\left(h/R_{col}\right)$, in which $h$ is the distance
of the sphere from the wall.
\begin{figure}
\includegraphics[width=7cm,height=8cm,angle=270]{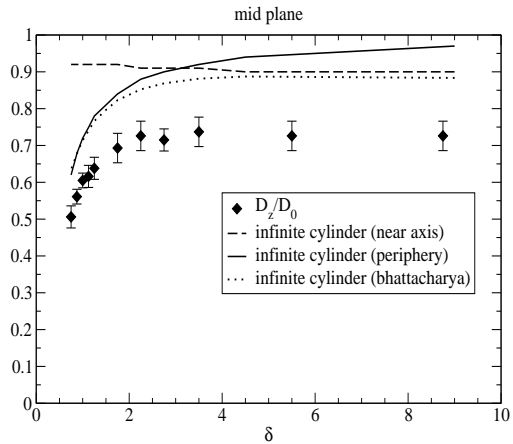}
\caption{\label{fig:theory1}\footnotesize{ Axial diffusion coefficient in the
    mid-plane. Simulations (diamond symbols) are compared with:
    i) prediction for an infinitely long cylinder, valid near the axis (dashed line),
    ii) predictions for a flat wall, valid at the periphery of the cylinder (solid line), and 
    iii) predictions from numerical solutions of the Navier-Stokes equations in an infinitely long cylinder (dotted
    line). All these predictions disregard the role of the end-walls, so that
    the diffusion is globally overestimated relative to our simulation results.
    }}
\end{figure}

In Fig.~(\ref{fig:theory1}) we plot the simulation
data for $D_z$  and the theoretical predictions
for an infinitely long cylinder. We use eq.~(\ref{eq:axcylinder}) which
is valid near the cylinder axis (dashed line) and
eq.~(\ref{eq:parallel}) which is valid near the the cylinder walls
(solid line). The profile provided for the region nearby the
cylindrical wall is also in good agreement with the model in
\cite{falade88}. Although such equations have only a limited range of
validity, in Fig.~(\ref{fig:theory1}) they have been plotted along the
entire range of distances we are interested in. In the same picture we
also show
the diffusion obtained from the numerical solution of the Navier-Stokes
equations (dotted line) for a cylinder provided by
S. Bhattacharya (private communication), according
to the method described in \cite{bhattacharya10}. We see 
that the latter result agrees very well with eqs~(\ref{eq:axcylinder}) and
(\ref{eq:parallel}) in the appropriate limits. The
discrepancy with our simulation results, however, is quite large for all values
of $\delta$.  This comparison clearly shows that the end walls' effects are important
for all particle positions in the mid-plane of the cylinder both close to the axis, as well as in
proximity of the cylinder walls, in case the separation between the end-walls is of the order
of the cylinder diameter or less.

In the following we will attempt to find a simple model that can predict the combined
effects of all walls, at least with the dimensions of our simulation box. 

\begin{figure}
\includegraphics[width=5cm,height=5cm]{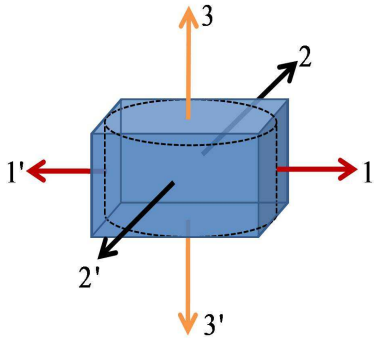}
\caption{\label{fig:parallelepipedum}\footnotesize{ Rectangular parallelepiped
circumscribing the cylinder. For a generic
    particle position, the indexes 1 and 1' refer to
    the walls perpendicular to the radial motion; 2 and 2' to the
    walls perpendicular to the azimuthal motion; 3 and 3' for the walls
    perpendicular to the axial motion.
    }}
\end{figure}

It has been shown \cite{happel74} that when the
particle is placed on the axis of a container and the size of the
particle is small compared to the width of the container, the axial
friction depends only weakly on the precise geometry, whether cylindrical or rectangular.
Inspired by this observation, we approximate the closed
cylinder by a rectangular parallelepiped circumscribed to it, as shown in Fig.~(\ref{fig:parallelepipedum}).
We then express the total correction factor $\Gamma$ for the friction
as a combination of single flat-wall corrections $\gamma$:
\begin{eqnarray}
\Gamma&=&
\gamma_1(d_1)+\gamma_{1'}(d_{1'})+\gamma_2(d_2)+\gamma_{2'}(d_{2'}) \nonumber \\
&& +\gamma_3(d_3)+\gamma_{3'}(d_{3'})-(n-1),
\label{eq:gamma_tot}
\end{eqnarray}
where $d_i$ refers to the distance of the particle from flat wall ``$i$'', as numbered
in Fig.~(\ref{fig:parallelepipedum}), $n$ is the number of
boundaries or walls
included in the model. In the parallelepiped model, $n=6$. We have to
subtract $n-1$ in order to avoid overcounting the bulk value;
this is a mere consequence of the definition of the wall correction factor. 

Before continuing with the closed cylinder, let us consider how our model performs for infinitely long systems, in particular for an infinitely long cylinder and for an infinitely long square duct. Our model in this case has just for flat walls, since we must remove the walls $3$ and $3'$ in
Fig.~(\ref{fig:parallelepipedum}); consequently $n=4$ in this case. For the cylinder we have 
 $\gamma_{axis}=1.132$, when $R_{cyl}=18R_{col}$; for the square duct we have
$\gamma_{sd}=1.118$ when the half-width of the square is equal to
$18R_{col}$
 \cite{happel74}
and for our model
we find $\Gamma=1.129$. All three numbers agree reasonably well.

\begin{figure}
\includegraphics[width=7cm,height=8cm,angle=270]{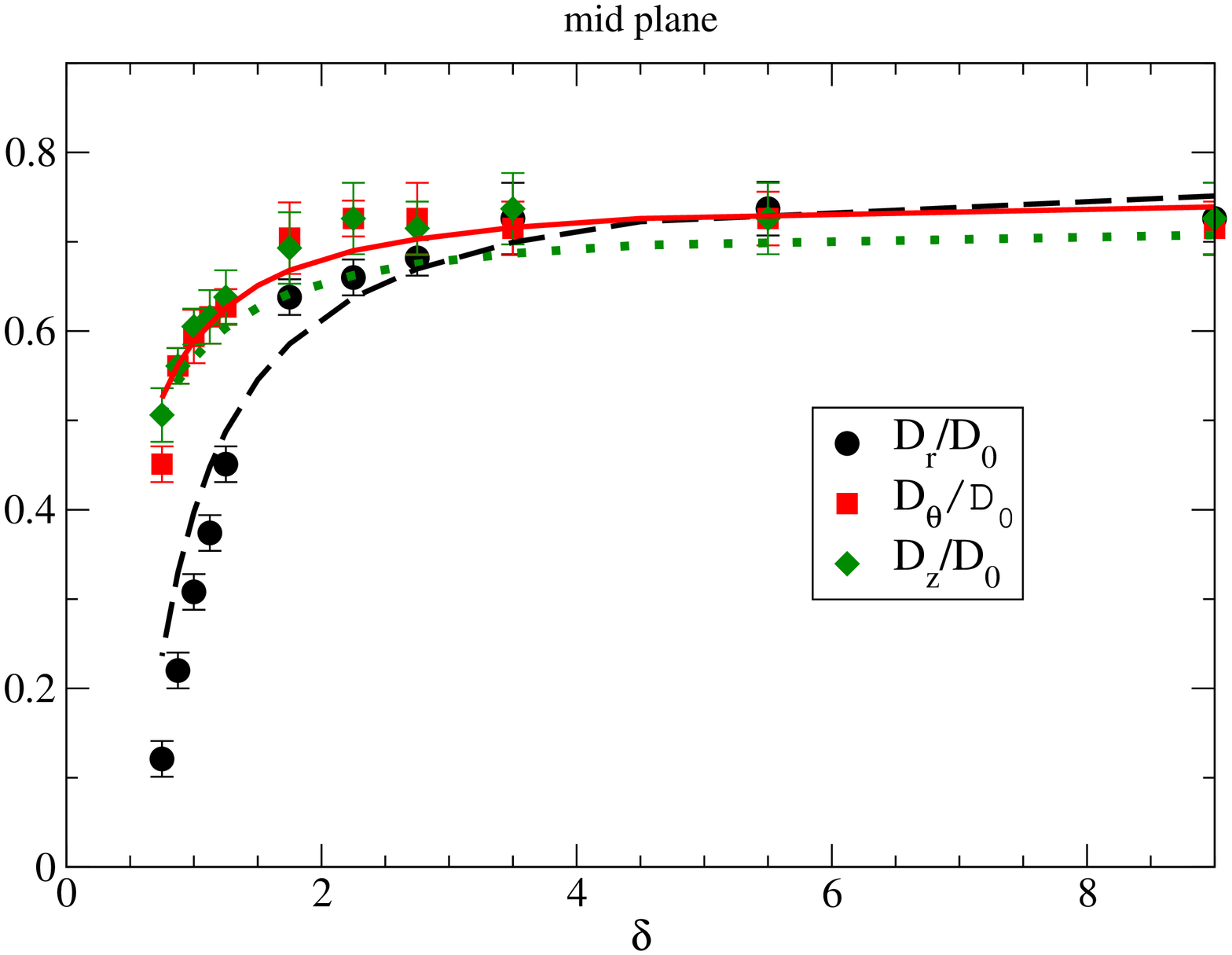}\\
\includegraphics[width=7cm,height=8cm,angle=270]{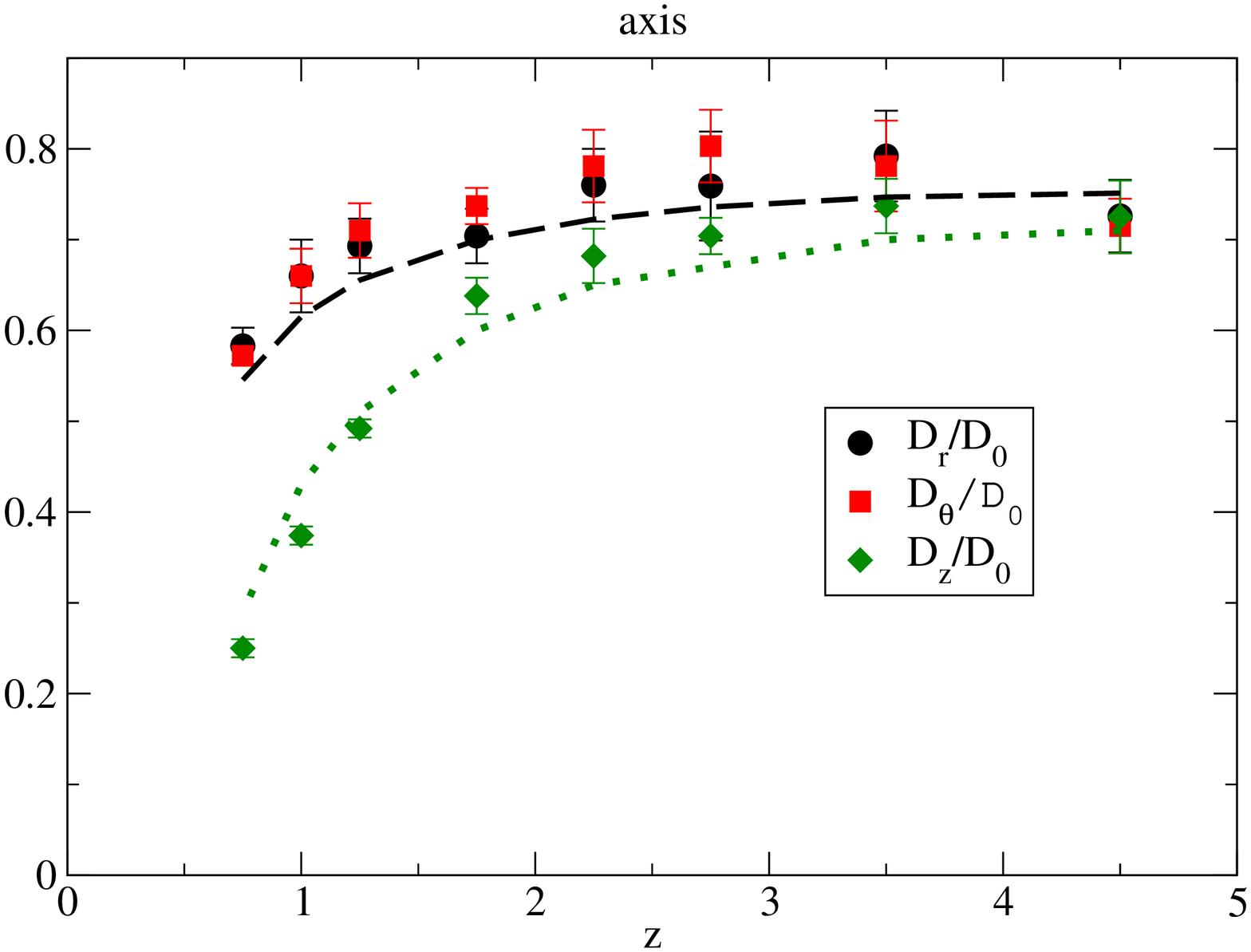}
\caption{\label{fig:theory2}\footnotesize{ Upper panel: diffusion in the
    mid-plane, where $\delta$ is the distance from the cylinder wall.
Bottom panel: diffusion along the axis, where $z$ is the distance from
the bottom wall. Symbols are for
    simulation data. Lines for the rectangular parallelepiped model: $D_r$ (dashed
    line), $D_\theta$ (solid line), $D_z$ (dotted line).
    }}
\end{figure}

Coming back to the case of the closed cavity, the motion along the
radial direction can be decomposed into a motion perpendicular to
 two walls and parallel with respect to the remaining four
walls. Similar conclusions can be drawn for the azimuthal and axial
motion.
So we can specialize the relation in (\ref{eq:gamma_tot}) in the
following manner:
\begin{eqnarray}
\Gamma_r&=&\gamma_{\perp,1}(d_1)+\gamma_{\perp,1'}(d_{1'})+
\gamma_{||,2}(d_2)+\gamma_{||,2'}(d_{2'}) \nonumber \\
&& +\gamma_{||,3}(d_3)+\gamma_{||,3'}(d_{3'})-5
\label{eq:Gamma1r} \\
\Gamma_{\theta}&=&\gamma_{||,1}(d_1)+\gamma_{||,1'}(d_{1'})+
\gamma_{\perp,2}(d_2)+\gamma_{\perp,2'}(d_{2'}) \nonumber \\
&& +\gamma_{||,3}(d_3)+\gamma_{||,3'}(d_{3'})-5
\label{eq:Gamma1theta} \\
\Gamma_z&=&\gamma_{||,1}(d_1)+\gamma_{||,1'}(d_{1'})+ \gamma_{||,2}(d_2)+\gamma_{||,2'}(d_{2'}) \nonumber \\
&& +\gamma_{\perp,3}(d_3)+\gamma_{\perp,3'}(d_{3'})-5
\label{eq:Gamma1z}
\end{eqnarray}
where the expression for the parallel ($||$) and perpendicular ($\perp$) corrections
are given by eqs (\ref{eq:parallel}) and (\ref{eq:perpendicular}), respectively.
In Fig.~(\ref{fig:theory2}) we compare these predictions with our simulation results;
the upper panel refers to the mid-plane, while
the bottom panel to the particle placed along the container axis.
The agreement is very good, which is rather surprising given the
simplicity of the model. Note however, that the radial and azimuthal diffusion coefficients predicted by our model are not exactly equal on the axis, as they should be.
Its success is possibly caused by a subtle cancellation of errors, which we will discuss
next.

The rectangular parallelepiped model is based upon a linear superposition of the
walls' effects. In this approximation, we disregard the influence of the curvature of the vertical walls and that of the corners obtained by the intersection of the cylindrical wall and the top and bottom walls. For a particle in the mid-plane and close to the vertical wall, ignoring the effect of the curvature, the presence of the corners obtained by the intersection of the cylindrical wall and the top and bottom walls leads to larger frictions, so that their neglect leads to an underestimation of the friction and consequently to an overestimation of the diffusion coefficients. 

On the other hand, using the superposition model for the top and bottom walls per se leads to a slight overestimation of the perpendicular frictions and thus to slight underestimations of the corresponding diffusions. For the slit geometry this may be seen in Fig.(4) in ref.\cite{bhattacharya05}. The magnitude of such effects depends on the
distance of the particle from the walls. For the present case, the parallel friction for a slit geometry is 
quite well represented by the linear superposition model, whereas the perpendicular friction is
overestimated.

For the configurations studied here the overestimation of
the friction perpendicular to the end walls is large enough to
compensate the underestimation of the total friction which is intrinsic to the
superposition model applied to a geometry with corners.

The above discussion can also be cast in the framework
of the method of reflections, where the role of walls is mimicked
through a set of images of the colloidal particle.
This set of images is used to enforce a zero velocity boundary condition
for the fluid near the planar surfaces. For a real parallelpid the superposition model only takes
into account the six images due to the single reflections on each of the
sides of the parallelepiped.
However, every time a new image is introduced (to treat a new wall)
the fluid velocity profile on the existing walls is modified. This is then corrected,
usually in an iterative fashion. Corner effects arise from the reflection
of each image particle with respect to two perpendicular walls.
In the slit geometry, instead, successive corrections to the linear
superposition, comes from the multiple reflections of the images,
similarly to those that appear in a barber shop.

\begin{figure}
\includegraphics[width=7cm,height=8cm,angle=270]{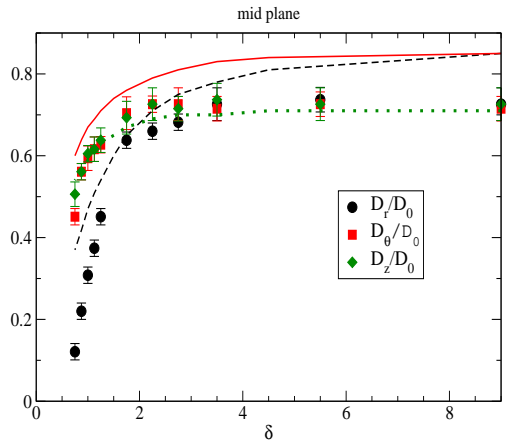}
\caption{\label{fig:bhatt}\footnotesize{ Comparison between
    simulation data and the superposition model applied to a
    combination of an infinitely long cylinder and two flat
    walls. Symbols are for
    simulation data. Lines for the rectangular parallelepiped model: $D_r$ (dashed
    line), $D_\theta$ (solid line), $D_z$ (dotted line).
    }}
\end{figure}

We have also tried a superposition of the friction expressions for an infinitely long
cylinder (private communication from Bhattacharya) and two planar walls. In this case the agreement with the
simulation data is less good, especially for the radial and azimuthal
diffusion  as shown in fig.\ref{fig:bhatt}
This result shows that the
 superposition model is quite sensitive to the details of the surfaces
 we apply it, and  therefore it should be applied with care.

\section{\label{sec6}Conclusions}
We have studied, by means of particle based simulations, the diffusion of a
spherical particle inside a microcavity in the shape of a
squat closed cylinder. We have analyzed the diffusion in the middle plane as
a function of the particle-cylinder separation $\delta$ and along the
axis of the cylinder as a function of the particle-end wall separation
$z$. For our squat cylinder we find that when the sphere is near the center of the cylinder
the diffusion is the same along the radial, azimuthal and axial
direction and it is significantly lower than in the bulk. 
In the mid-plane, when the particle is shifted
towards the cylinder wall, the diffusion decreases more in the radial
direction than in the azimuthal and axial ones. The last two are very
similar to each other. When  the particle is shifted along the axis we
find that the diffusion decreases much more along the axial direction than
perpendicular to it. In general we can say that the diffusion in the
direction mostly parallel to the closest
confining wall, decreases slower than in the direction perpendicular to it. 

We have compared our simulations with experimental results for a $\mu
m$ sized particle in a cylindrical microcavity, finding good agreement. We also find good 
agreement with the theoretical predictions for a closed cylinder reported in \cite{lecoq07}.

Finally we have shown that the simplest approximation to capture confinement effects,
namely one which uses the result for an infinitely long cylinder for the axial diffusion, does not work.
Neither does treating the closest wall as a flat wall when the particle is at the periphery of the box.
The end-wall effects appear very
important for all positions of the sphere inside the squat cylinder. We have investigated a
simplified model in which the effect of all walls is approximated as a superposition of the effects of
individual walls of a rectangular parallelepiped circumscribing the cylinder. 
This model works well for diffusion coefficients both in the
mid-plane and along the axis, but it owes its success to a fortunate cancellation of errors.
\begin{acknowledgments}
We thank Sukalyan Bhattacharya
for fruitful discussions, suggestions and kindly providing the numerical data to compare
and validate our simulation results.
\end{acknowledgments}


\end{document}